\documentclass[11pt,draftcls,onecolumn] {IEEEtran}

\ifCLASSINFOpdf
\else
\fi
\usepackage[cmex10]{amsmath}
\usepackage{algorithmic}

\usepackage{bbm}
\usepackage{array}

\usepackage[tight,footnotesize]{subfigure}

\usepackage{graphicx,cite}
\usepackage{latexsym}
\usepackage{multirow}
\usepackage{amssymb}

\usepackage{url}

\begin{document}
%

\title{Pilotless Recovery of Clipped OFDM\\ Signals by Compressive Sensing over\\ Reliable Data Carriers}

%

\author{\IEEEauthorblockN{Ebrahim B. Al-Safadi}
\IEEEauthorblockA{Inteltec EPC\\
Riyadh, Saudi Arabia\\
Email: esafadi@saudi-inteltec.com} \\\and 
\IEEEauthorblockN{Tareq Y.
Al-Naffouri}
\IEEEauthorblockA{Department of Electrical Engineering\\
King Abdullah University of Science and Technology\\
Thuwal, Saudi Arabia\\
Email: tareq.alnaffouri@kaust.edu.sa} }


%


\maketitle

\begin{abstract}

In this paper we propose a novel form of clipping mitigation in OFDM
using compressive sensing that completely avoids tone reservation
and hence rate loss for this purpose. The method builds on selecting
the most reliable perturbations from the constellation lattice upon
decoding at the receiver, and performs compressive sensing over
these observations in order to completely recover the temporally
sparse nonlinear distortion. As such, the method provides a unique
practical solution to the problem of initial erroneous decoding
decisions in iterative ML methods, offering both the ability to
augment these techniques and to solely recover the distorted signal
in one shot.
\end{abstract}


%
\IEEEpeerreviewmaketitle

\section{Introduction}

\IEEEPARstart{M}{ulticarrier} signalling schemes such as Orthogonal
Frequency Division Multiplexing (OFDM) have an inherent sensitivity
to nonlinear distortion at all stages of the transmission process.
To obtain information about the nonlinear temporal distortion in an
OFDM signal, the majority of receiver-based mitigation techniques
begin with observing the deviation of the equalized frequency domain
variables from the discrete symbol constellation. As useful as this
may be, a valid inconsistency is always persistently present. After
all, it is the position of those very symbols in the frequency
domain that ultimately entitle our decoding decisions, and should
any of those symbols be perturbed outside their correct decision
boundaries by nonlinear distortion, it will always be the case that
any further reliance on these erroneous measurements might be
resistent to further correction. Furthermore, refraining from using
part of the deviations in recovering the distortion reduces the
effectiveness of the mitigating algorithm.

Our major contributions are then to first suggest algorithms that
can use a subset of the deviations in the frequency domain to dually
avoid erroneous decisions and recover from the distortion with no
theoretical sacrifice of given information and thus performance, and
secondly to tailer the input model to these algorithms by selecting
the most appropriate set of observations using a simplified
procedure that models an actual Bayesian reliability measure.
Although many scenarios and modifications apply to the methods
herein, due to the limited space and the ongoing development of the
presented concepts, we will restrict our discussion to mitigating
distortion caused by clipping at the transmitter, and delay more
elaborate applications to a further treatment.


Unless otherwise noted, frequency domain variables will be
represented by uppercase italic letters while lower case letters
will be reserved for time domain variables. The lower index in
$\mathcal{X}_{i}$ will denote the $i^{th}$ constellation point
amongst an M-ary alphabet $\mathcal{X}$ while $A_{i}(k)$ will be
used for the $k^{th}$ scalar coefficient of the the $i^{th}$ column
vector $A_{i}$ of matrix $\textbf{A}$. Furthermore, $\langle
X(k)\rangle$ will denote a hard decoding operation which maps $X(k)$
back into $\mathcal{X}$. The standard notation of $x_{i:N}$ will be
be used for the $i^{th}$ order statistic in a sample of $N$ random
variables of a common probability density function \cite{David}.
Finally, we use $\mathbb{F}$ for Cumulative Distribution Functions
(CDF) and $\textbf{F}$ for unitary Fourier matrices.

%
\section{Transmission and Clipping Model}

In an OFDM system, Serially incoming bits are mapped into an
\textit{M}-ary QAM alphabet
$\{\mathcal{X}_{0},\mathcal{X}_{1},\ldots,\mathcal{X}_{M-1}\}$ and
concatenated to form an $N$ dimensional data vector $X=[X(0)
X(1)\cdots X(N-1)]^{T}$. The time-domain signal is obtained by an
IFFT operation so that $x=\textbf{F}^{H}X$ where
\begin{equation*}
F_{k}(\ell)=N^{-1/2}\, e^{-j2\pi k\ell/LN},\quad k,\ell\in
{0,1,\ldots,LN-1}.
\end{equation*}
and $L$ is an oversampling factor. Since $x$ has a high peak to
average power ratio (PAPR), the digital samples are subject to a
magnitude limiter which saturates its operands to a value of
$\gamma$, and hence instead of feeding $x$ to the power amplifier,
we feed $\bar{x}$ where
\begin{eqnarray} \bar{x}(i) =
\begin{cases}
\gamma e^{\;j\theta_{x(i)}}\;  & \mbox{if} \;|x(i)|>\gamma, \\
\,\,\,\,x(i)\; & \mbox{otherwise}
\end{cases}
\end{eqnarray}
where $\theta_{x(i)}$ is the phase of $x(i)$. This soft limiting
operation can be conveniently thought of as adding a peak-reducing
signal $c$ to $x$ whereby its low-PAPR counterpart $\bar{x}=x+c$ is
transmitted instead, and whereby $x$ can be re-generated at the
receiver by estimating $c$. What's more, by setting a typical
clipping threshold $\gamma$ on $x$, $c$ is controllably sparse in
time by the impulsive nature of $x$, and dense in frequency by the
uncertainty principle. We will denote its temporal support by
$\mathcal{I}_{c}=\{n:c(n)\neq 0\}$ and always maintain the practical
assumption that $|\mathcal{I}_{c}|\ll N$.

In the frequency domain, this translates to transmitting
$\bar{X}=X+C$, with complex coefficients that are now randomly
pre-perturbed from the lattice $\mathcal{X}$, followed by additional
random post-perturbations by the channel
$\textbf{H}\!\!=\!\!\textbf{F}^{H}\Lambda \textbf{F}$ and additive
noise samples $Z\sim\mathcal{CN}(0,\sigma_{Z}I_{N\times N})$ at the
receiver, where the circulant channel $\textbf{H}$ has been
decomposed as such by virtue of the added cyclic prefix in OFDM
signalling. At the receiver, this reads
\begin{eqnarray}
Y=\Lambda \,\bar{X}+Z,
\end{eqnarray}
where we will make the practical assumption that the channel
coefficients are known on its side. Consequently, $\bar{X}$ can be
directly recovered scalar-wise from $Y$, i.e.
\begin{eqnarray}\label{X_bar_estimate}
\nonumber\hat{\bar{X}}(k)&=&\Lambda^{-1}_{k}(k)Y(k)\\
&=&X(k)+C(k)+\Lambda_{k}^{-1}(k)Z(k).
\end{eqnarray}
Let $D(k)\!\triangleq \!C(k)+\Lambda_{k}^{-1}(k)Z(k)$ denote the
general distortion on the frequency domain sample
$X(k)$.\footnote{$D(k)$ is a random variable with a PDF that is a
function of $\gamma$, $\Lambda^{-1}_{k}(k)$, $\sigma_{Z}$, and a
compound distribution $f_{C(k)}$ which must be conditioned and then
marginalized over the random support $\mathcal{I}_{c}$. We avoid
presenting its derivation and justifying its proximity to a Gaussian
in this paper due to lack of space, and directly treat it as a
circularly symmetric variable with parameter $\sigma_{D(k)}$. For
the same reason, we also express functions compactly in terms of
$f_{D(k)}(\cdot)$ by manipulating its argument only.} A naive ML
decoder will now simply map $\hat{\bar{X}}(k)$ to the nearest
constellation point $\mathcal{X}_{i^{*}}$ to recover $X(k)$, where
$i^{*}(k)\triangleq
\arg\min_{i}|\hat{\bar{X}}(k)-\mathcal{X}_{i}(k)|$, treating the
clipping distortion as additive noise. Although such a hard-decoding
scheme is very efficient at high SNR in the classical AWGN scenario,
the clipping scenario, however, introduces another
$\gamma$-dependent source of perturbation which is immune to any
increase in SNR.


An intelligent ML decoder will hence have to iteratively update its
decisions in the frequency domain based on the resulting waveforms
in the time domain. Unfortunately, such a method will suffer from
error propagation since a single faulty decision in frequency will
generate a faulty estimate of $c$ in time which will be used to
update the frequency perturbations in the next iteration and so on.

A direct countermeasure would be to refrain from using the tones at
which the perturbations $D(k)$ are large and hence unreliable
\cite{Quasi}. Although this should eliminate false positives in the
time domain, the economy in tone usage severely limits the
improvement offered by such an approach.

Alternatively, CS seems to be a very sensible solution to this
problem. A partial observation of the frequency content of a sparse
signal in the time domain is sufficient to recover $c$ and hence $C$
in one shot. This would certainly get around the problem of
unreliable perturbations as CS algorithms can be totally blind to
them and still offer near optimal signal reconstruction under mild
conditions.


Fortunately, unlike our previous approach \cite{Safadi} of reserving
a sufficient number of tones at the transmitter to recover $c$, and
consequently reducing the transmission rate, we do not require any
tone reservation in this method, and are completely free to choose
any subset $\Omega_{m}$ from the $N$ data-carrying tones in order to
reconstruct $c$ at the receiver. This freedom of choice opens up
many possibilities in how to select particular adaptive subsets to
optimize the CS performance as will be thoroughly discussed later
on.

\section{Development of Compressive Sensing Models with No Tone Reservation}

With the addition of $C$ to the data vector $X$, we suspect that a
part of the data samples $X(k)$ will be severely perturbed to fall
out of their corresponding decision regions $\mathcal{A}_{X(k)}$.
Denote by $\Omega_{T}=\{k:\langle X(k)+C(k)\rangle=X(k)\}$ the
subset of data tones in $\Omega$ in which the perturbations are not
severe (i.e. do not cause crossing a decision boundary). At these
locations, the equality in $\langle \bar{X}(k)\rangle=X(k)$ is true
and hence
$C_{\Omega_{T}}=\bar{X}_{\Omega_{T}}-\langle\bar{X}_{\Omega_{T}}\rangle$
at the transmitter. More generally,
\begin{eqnarray}
C=\textbf{S}_{\Omega_{T}}\left(\bar{X}-\langle
\bar{X}\rangle\right)+\textbf{S}_{\bar{\Omega}_{T}}\left(\bar{X}-X\right)
\end{eqnarray}
where $\textbf{S}_{\Omega_{T}}$ is an $N\!\times \!N$ diagonal and
binary selection matrix, with $|\Omega_{T}|$ ones along its diagonal
that extract the locations in the vector $\bar{X}-\langle
\bar{X}\rangle$ according to the tone set $\Omega_{T}$ while nulling
the others, and $\textbf{S}_{\bar{\Omega}_{T}}$ is its complement
such that
$\textbf{S}_{\Omega_{T}}\textbf{S}_{\bar{\Omega}_{T}}=\textbf{0}_{N\times
N}$. Practically speaking, $\Omega_{T}$ constitutes the bigger part
of the general tone set $\Omega$, with a probability of occupying at
least $100\alpha\%$ of $\Omega$ equal to $\Pr(|\Omega_{T}|>\alpha
N)\approx\sum_{\ell=0}^{N(1-\alpha)}{N\choose\ell}P_{e}^{\ell}(1-P_{e})^{N-\ell}$
for large constellations, where
$P_{e}=2Q\left(\frac{d_{\min}}{2\sigma_{D}}\right)$. An essential
part of OFDM signal recovery obviously constitutes finding this set,
and correcting the distortion over $\bar{\Omega}_{T}$ to finally
reach $\Omega_{T}=\Omega$.

Upon demodulation and decoding at the receiver, we are left with an
estimate $\hat{\bar{X}}$ of the distorted data vector given in
(\ref{X_bar_estimate}) along with its associated decoded vector
$\langle\hat{\bar{X}}\rangle\in\mathcal{X}^{N}$. Taking the
difference yields
\begin{eqnarray*}
\hat{\bar{X}}-\langle\hat{\bar{X}}\rangle&=&X+D-\langle X+D\rangle\\
&=&X+D-(\textbf{S}_{\Omega_{T}}X+\textbf{S}_{\bar{\Omega}_{T}}E)
\end{eqnarray*}
where $\Omega_{T}$ now indexes the locations where $X(k)+D(k)$
remains within the correct ML decision region and $E$ represents the
error vector resulting from incorrect decoding decisions at
$\bar{\Omega}_{T}$. Multiplying both sides by
$\textbf{S}_{\Omega_{T}}$ leaves us with
\begin{eqnarray}
\nonumber\textbf{S}_{\Omega_{T}}(\hat{\bar{X}}-\langle\hat{\bar{X}}\rangle)&=&\textbf{S}_{\Omega_{T}}X+\textbf{S}_{\Omega_{T}}D-\textbf{S}_{\Omega_{T}}(\textbf{S}_{\Omega_{T}}X+\textbf{S}_{\bar{\Omega}_{T}}E)\\
\nonumber &=&\textbf{S}_{\Omega_{T}}X+\textbf{S}_{\Omega_{T}}D-\textbf{S}_{\Omega_{T}}X+\textbf{0}_{N\times 1}\\
\nonumber &=&\textbf{S}_{\Omega_{T}}D\\
&=&\textbf{S}_{\Omega_{T}}\textbf{F}c+\textbf{S}_{\Omega_{T}}\Lambda^{-1}Z
\end{eqnarray}
where we have used the fact that
$\textbf{S}_{\Omega_{T}}^{n}=\textbf{S}_{\Omega_{T}}$ for any
positive integer $n$, and redundantly used
$\textbf{S}_{\bar{\Omega}_{T}}$ on $E$ to show that
$\textbf{S}_{\Omega_{T}}E=\textbf{S}_{\Omega_{T}}\textbf{S}_{\bar{\Omega}_{T}}E=\textbf{0}_{N\times
1}$. Note, however, that we do not require all of $\Omega_{T}$ to
recover $c$, for obviously there would be no need for any recovery
algorithm if we knew $\Omega_{T}$. Rather, we only require an
arbitrary subset $\Omega_{m}\subseteq\Omega_{T}\subseteq\Omega$ of
cardinality $|\Omega_{m}|<|\Omega_{T}|$ to correctly recover $c$ by
CS. As a result, we can replace the equation above with
\begin{eqnarray*}
\textbf{S}_{\Omega_{m}}(\hat{\bar{X}}-\langle\hat{\bar{X}}\rangle)&=&\textbf{S}_{\Omega_{m}}\textbf{F}c+\textbf{S}_{\Omega_{m}}\Lambda^{-1}Z\\
&=&\Psi c+Z'
\end{eqnarray*}
where $\Psi\triangleq\textbf{S}_{\Omega_{m}}\textbf{F}$,
$Z'\triangleq\textbf{S}_{\Omega_{m}}\Lambda^{-1}Z$, and where we
further let
$Y'\triangleq\textbf{S}_{\Omega_{m}}(\hat{\bar{X}}-\langle\hat{\bar{X}}\rangle)$
denote the observation vector of the differences over the tones in
$\Omega_{m}$, nulled at the discarded measurements. This leads us
to the lossless-rate CS model
\begin{eqnarray}\label{lossless_CS_model}
Y'_{\Omega_{m}}=\Psi_{\Omega_{m}} c+Z_{\Omega_{m}}'.
\end{eqnarray}
where $Y'_{\Omega_{m}}$ is the $|\Omega_{m}|$-dimensional vector
collecting the nonzero coefficients in $Y'$. Such a generic model
can now be processed for $c$ using any compressive sensing
technique, be it convex programming, greedy pursuit, or iterative
thresholding, and a very flexible region for tradeoff exists in
regard to performance and complexity. In any case, our subsequent
objective is to scrutinize the general conditioning of the model
itself by supplying our most reliable observations to the generic CS
algorithm.

\section{Cherry Picking $\Omega_{m}$}\label{tone_selection_criteria}

An essential question now is how one is to select among the $N
\choose m$ possible constructions of $\Omega_{m}$. A general
strategy of CS techniques is to select these $m$ tones randomly for
near-optimum performance. Although possible in this scenario, such a
strategy neglects the fact that our observations vary in their
credibility and attest to wether they represent true
frequency-domain measurements of $C$ or not since our assumption
that $\hat{\bar{X}}(k)-\langle\hat{\bar{X}}(k)\rangle=D(k)$ is
probabilistic. Furthermore, it neglects the fact that the estimation
signal-to-noise-ratio $\mathbb{E}[\|\Psi_{\Omega_{m}}
c\|_{2}^{2}]/\mathbb{E}[\|Z_{\Omega_{m}}'\|_{2}^{2}]$ also varies
with the channel gains $\{\Lambda_{k}(k)\}_{k\in\Omega_{m}}$, and
that knowledge of these gains has an effect on our reliability
estimates.\footnote{We will refer to this ratio as the
clipper-to-noise ratio (CNR) in order not to confuse it with the
transmission model's SNR,
$\mathbb{E}[\|\Lambda\bar{x}\|_{2}^{2}]/\mathbb{E}[\|z\|_{2}^{2}]$.}
With the receiver risking faulty decisions, it must
devise a procedure to select the most reliable set of observations
in which to sense over. This could be done based on the relative
posterior probability of $D(k)$ equalling
$\hat{\bar{X}}(k)-\langle\hat{\bar{X}}(k)\rangle$ to the probability
of it equaling some other difference vector
$\hat{\bar{X}}(k)-\mathcal{X}_{i,i\neq i^{*}}$. More precisely, let
\begin{eqnarray}
\nonumber\mathfrak{R}(k)&=&\log\frac{\Pr(\langle\hat{\bar{X}}(k)\rangle=X(k)|\hat{\bar{X}}(k))}{\Pr(\langle\hat{\bar{X}}(k)\rangle=\mathcal{X}_{\textmd{NN}}(k)|\hat{\bar{X}}(k))}\\
&=&\log\frac{\Pr(D(k)=\hat{\bar{X}}(k)-\langle\hat{\bar{X}}(k)\rangle)}{\Pr(D(k)=\hat{\bar{X}}(k)-\mathcal{X}_{\textmd{NN}}(k))}\label{reliability_NN}
\end{eqnarray}
\begin{figure}
\centering
\includegraphics[width=3.0in]{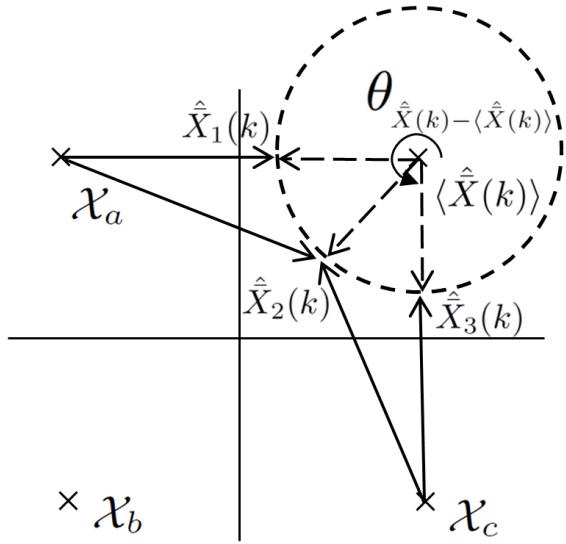}
\caption{Variation of the reliability of observation
$\hat{\bar{X}}(k)-\langle\hat{\bar{X}}(k)\rangle$ as the relative
distances between it and the other constellation points changes with
$\theta_{\hat{\bar{X}}(k)-\langle\hat{\bar{X}}(k)\rangle}$.}
\label{NTR_Illustration_CNF2}
\end{figure}
define the reliability in decoding $\hat{\bar{X}}(k)$ to the closest
constellation point relative to decoding to the nearest neighbor
$\mathcal{X}_{\textmd{NN}}(k)$. The minimum certainty occurs at the
boundary of the decision region and attains
$\mathfrak{R}_{\min}(k)=0$. At such tones, we would be highly
skeptical of whether
$D(k)=\hat{\bar{X}}(k)-\langle\hat{\bar{X}}(k)\rangle$ or
$D(k)=\hat{\bar{X}}(k)-\mathcal{X}_{\textmd{NN}}(k)$, and would
hence be supplying a plausibly false measurement to the CS
algorithm. Instead, assume we only chose tones where
$|\hat{\bar{X}}(k)-\langle\hat{\bar{X}}(k)\rangle|$ were confined to
a disk of radius $r$. In such a case, the minimum reliability would
increase to
$\mathfrak{R}_{\min}(k)=\log\frac{f_{D(k)}(r)}{f_{D(k)}(d_{\min}-r)}$
in case of the nearest neighbor $\mathcal{X}_{\textmd{NN}}$, and to
$\mathfrak{R}(k)=\log\frac{f_{D(k)}(r)}{f_{D(k)}\left(\sqrt{2}d_{\min}-r\right)}$
for the next nearest neighbor $\mathcal{X}_{\textmd{NNN}}$ measured
in the direction of a decision region's corner. The reliability of a
measurement at each tone is then a function $\mathfrak{R}(k)$ that
maps a 3-tuple
$(|\hat{\bar{X}}(k)-\langle\hat{\bar{X}}(k)\rangle|,\theta_{\hat{\bar{X}}(k)-\langle\hat{\bar{X}}(k)\rangle},\Lambda_{k}^{-1}(k))$
into $\mathbb{R}_{0}^{+}$. Fig. \ref{NTR_Illustration_CNF2}
illustrates this concept such that, for example, even though
$|\hat{\bar{X}}_{1}(k)-\langle\hat{\bar{X}}(k)\rangle|=|\hat{\bar{X}}_{2}(k)-\langle\hat{\bar{X}}(k)\rangle|$,
we have
\begin{eqnarray*}
\frac{|\hat{\bar{X}}_{1}(k)-\langle\hat{\bar{X}}(k)\rangle|}{|\hat{\bar{X}}_{1}(k)-\mathcal{X}_{a}|}>\frac{|\hat{\bar{X}}_{2}(k)-\langle\hat{\bar{X}}(k)\rangle|}{|\hat{\bar{X}}_{2}(k)-\mathcal{X}_{a}|}
\end{eqnarray*}
and so the reliability of assuming
$D_{2}(k)=\hat{\bar{X}}_{2}(k)-\langle\hat{\bar{X}}(k)\rangle$ is
higher than the reliability of assuming
$D_{1}(k)=\hat{\bar{X}}_{1}(k)-\langle\hat{\bar{X}}(k)\rangle$,
although
$f_{D(k)}(\hat{\bar{X}}_{1}(k)-\langle\hat{\bar{X}}(k)\rangle)=f_{D(k)}(\hat{\bar{X}}_{2}(k)-\langle\hat{\bar{X}}(k)\rangle)$
by the circular symmetry assumption on $D(k)$. Ultimately, we would
choose our measurements according to the tones associated with the
highest $m$ reliability outputs, i.e.
\begin{eqnarray}\label{ordered_reliability}
\Omega_{m}\triangleq
\arg\left\{\mathfrak{R}_{i:N}\right\}_{i=N-m+1}^{N}.
\end{eqnarray}
Luckily, the locations of these tones are random and hence such a
selection also preserves the near-optimality selection of tones for
generic CS performance.

\subsection{Bayesian Reliability}\label{Defining_Reliable_Regions}

Using the reasoning based on the probability $\Pr(\langle
\hat{\bar{X}}(k)\rangle=X(k)|\hat{\bar{X}}(k))$, an exact expression
for the reliability could be a direct generalization of
(\ref{reliability_NN}), namely,
\begin{eqnarray}\label{reliability_exact}
\mathfrak{R}(k)=\log\frac{f_{D(k)}(\hat{\bar{X}}(k)-\langle
\hat{\bar{X}}(k)\rangle)}{\mathfrak{R}_{\min}\sum_{^{i=0}_{i\neq
i^{*}}}^{M-1}f_{D(k)}(\hat{\bar{X}}(k)-\mathcal{X}_{i}(k))}
\end{eqnarray}
where the constant $\mathfrak{R}_{\min}$ is inserted to compensate
for the rare worst case scenarios and preserve $\mathfrak{R}(k)\geq
0$. For example, $\mathfrak{R}_{\min}=1/3$ would be sufficient for
the case when $\hat{\bar{X}}(k)$ falls on the center point between
four constellation points. Unfortunately, this pursuit for exact
reliability computation is inefficient. Even if we truncate the
summation in (\ref{reliability_exact}) to the nearest neighbors, the
method would still require repeating redundant evaluations of
$f_{D(k)}(\cdot)$. What is required is then a method that could
approximate $\mathfrak{R}(k)$ based solely on the observation
$\hat{\bar{X}}(k)-\langle \hat{\bar{X}}(k)\rangle$ with no reference
to any other constellation point $\mathcal{X}_{i}$.

\subsection{Practical Geometric-Based Reliability Computation}\label{practical_reliability}

The competitive constellation points can be accounted for by
considering the magnitude and phase of our observation against the
location of $\langle \hat{\bar{X}}(k)\rangle$ within the
constellation plane. For example, an observation with $\langle
\hat{\bar{X}}(k)\rangle$ being a midpoint in a large rectangular
constellation will have a higher reliability if its phase
$\theta_{\hat{\bar{X}}(k)-\langle \hat{\bar{X}}(k)\rangle}$ were
along $\left\{\frac{\pi}{4}+\frac{\pi}{2}i,\,i=0,1,2,3\right\}$,
compared to an observation with the same magnitude pointing in a
different direction, which ultimately reaches a minimum reliability
at phases $\left\{\frac{\pi}{2}i,\,i=0,1,2,3\right\}$. Therefore let
\begin{eqnarray}\label{square_reliability}
\mathfrak{R}^{|\cdot|,\theta}(k)=f_{D(k)}\left(\hat{\bar{X}}(k)-\langle
\hat{\bar{X}}(k)\rangle\right)g\left(\theta_{\hat{\bar{X}}(k)-\langle
\hat{\bar{X}}(k)\rangle}\right)
\end{eqnarray}
define a reliability function which is computed based on the
magnitude and phase of the respective $k^{th}$ coefficient alone. A
general function which was found to very closely match the exact
reliability outcome (\ref{reliability_exact}) for inner
constellation points is
\begin{eqnarray}
g\left(\theta_{\hat{\bar{X}}(k)-\langle
\hat{\bar{X}}(k)\rangle}\right)=\frac{\alpha}{\alpha+\beta}+\frac{\beta}{\alpha+\beta}\cos\left(4\theta_{\hat{\bar{X}}(k)-\langle
\hat{\bar{X}}(k)\rangle}+\pi\right)
\end{eqnarray}
where $\alpha>\beta>0$. Furthermore, the aim is to also make
$g(\cdot)$ magnitude dependent so that its profile supported by
$[0,2\pi]$ will be increasingly tapered along
$\left\{\frac{\pi}{4}+\frac{\pi}{2}i,\,i=0,1,2,3\right\}$ relative
to $\left\{\frac{\pi}{4}+\frac{\pi}{2}i,\,i=0,1,2,3\right\}$ as the
magnitude $|\hat{\bar{X}}-\langle \hat{\bar{X}}(k)\rangle|$
increases, compared to a fully isotropic profile at vanishingly
small magnitudes. By linearly mapping $\alpha/(\alpha+\beta)\in
[1/2,1]$ to $|\hat{\bar{X}}-\langle \hat{\bar{X}}(k)\rangle|\in
[0,d_{\min}]$ we finally obtain
\begin{eqnarray}\label{g_function}
\nonumber g^{|\cdot|,\theta}\left(\theta_{\hat{\bar{X}}(k)-\langle
\hat{\bar{X}}(k)\rangle}\right)=\frac{\sqrt{2}d_{\min}-|\hat{\bar{X}}(k)-\langle
\hat{\bar{X}}(k)\rangle|}{\sqrt{2}d_{\min}}\\
+\frac{|\hat{\bar{X}}(k)-\langle
\hat{\bar{X}}(k)\rangle|}{\sqrt{2}d_{\min}}\cos\left(4\theta_{\hat{\bar{X}}(k)-\langle
\hat{\bar{X}}(k)\rangle}+\pi\right)
\end{eqnarray}
\begin{figure}
\centering
\includegraphics[width=3.5in]{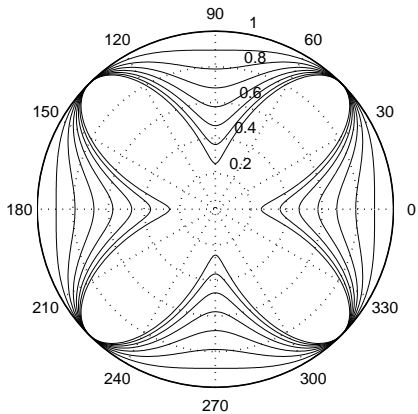}
\caption{Illustration of the phase penalty function
$g^{|\cdot|,\theta}\left(\theta_{\hat{\bar{X}}(k)-\langle
\hat{\bar{X}}(k)\rangle}\right)$ expressed in (\ref{g_function}).
The function is normalized, and therefore the outer circle-shaped
curves actually correspond to the smallest magnitudes, and become
more tapered as $|\hat{\bar{X}}(k)-\langle \hat{\bar{X}}(k)\rangle|$
increases.} \label{Phase_Illustration}
\end{figure}
\!which is portrayed in Fig. \ref{Phase_Illustration} for different
magnitudes. The last approximation we wish to mention is the simple
magnitude-based function
\begin{eqnarray}\label{mag_reliability}
\mathfrak{R}^{|\cdot|}(k)=f_{D(k)}\left(\hat{\bar{X}}(k)-\langle
\hat{\bar{X}}(k)\rangle\right)
\end{eqnarray}
which is completely blind to the other constellation points.
Nonetheless, for small $\sigma_{D}^{2}$ this approximation is very
efficient, especially for inner points in large constellations. Once
the type of function is set and the vector $\mathfrak{R}$ is
computed, we can directly select $\Omega_{m}$ from
(\ref{ordered_reliability}), fix our model
(\ref{lossless_CS_model}), and proceed to recovering $c$ by CS.

To be sure, we used two different schemes of CS to recover $c$ from
the developed CS model in (\ref{lossless_CS_model}), one from the
convex relaxation group and the other from greedy pursuit methods.
More specifically, the first is a weighted and phase-augmented LASSO
\cite{Tibshirani} we refer to as WPAL \cite{Safadi2}, which is a
data aided modification of the standard LASSO that incorporates data
in the time domain to improve distortion recovery, and can be
defined as
\begin{eqnarray}\label{LASSO}
\hat{c}=\arg_{c}\min
\||\textbf{F}^{H}\hat{\bar{X}}-\gamma|^{T}c\|_{1}
\,\,\textmd{s.t.}\,\, \|Y'_{\Omega_{m}}-\tilde{\Psi}_{\Omega_{m}}
c\|_{2}^{2}<\epsilon
\end{eqnarray}
for some noise-dependent parameter $\epsilon$. The other technique
is the Bayesian Matching Pursuit (BMP) by Schniter et al.
\cite{Schniter2} chosen for its superior performance and efficiency
when a relatively large amount of measurements is available to it, a
luxury we can actually enjoy in this work, unlike when pilot
reservation is used to construct the observation vector
$Y'_{\Omega_{m}}$ and an extreme economy in tones is enforced to
preserve data rate \cite{Safadi2}.
\section{Simulation Results}

The methods proposed in this paper were tested on an OFDM signal of
$64$ subcarriers drawn from a $16$-QAM constellation. The signal was
subject to a block-fading, frequency-selective Rayleigh channel
model with an SNR of $25$ dB per bit, and a severe clipping level
(defined as $10\log\gamma^{2}/\sigma_{x}^{2}$) of $2$ dB. No bit
loading (i.e. no variation of constellation size per carrier SNR),
diversity gain, or error control coding were considered. Special
packages for convex programming \cite{Boyd}, and greedy pursuit
\cite{Schniter2} were used to implement our CS algorithms.

\begin{figure}
\centering
\includegraphics[width=3.5in]{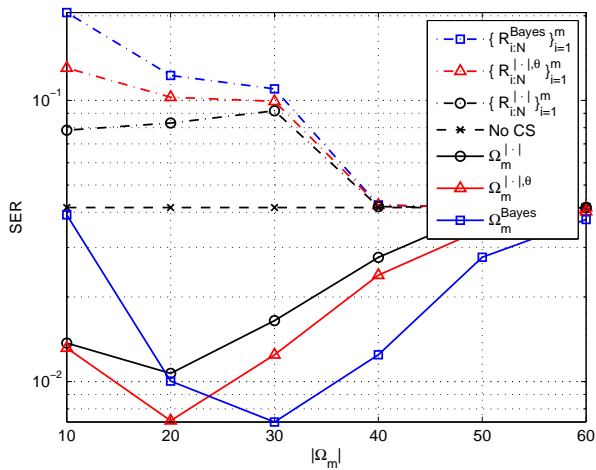}
\caption{SER vs. $|\Omega_{m}|$ for the various reliability
functions defined in (\ref{reliability_exact}),
(\ref{square_reliability}), and (\ref{mag_reliability}) and their
least reliable counterparts.} \label{Reliability_fig}
\end{figure}

Fig. \ref{Reliability_fig} shows the result of using WPAL
(\ref{LASSO}) with the proposed reliability criteria in
\ref{tone_selection_criteria} for choosing the measurement tone set
$\Omega_{m}$. We plotted the results against an increased number of
observed tones, such that, for instance, the most $10$ reliable
observations are used, compared to using the most $20$ reliable
observations, and so on. In doing so we expect a somewhat convex
behavior of the SER as a function of $|\Omega_{m}|$, since generally
the more observations we use the better the performance of CS
algorithms become (up to some typical saturation level), but then
due to the increased amount of erroneous observations supplied as
$|\Omega_{m}|$ increases, the performance eventually deteriorates.
The simulation results confirm this intuition, and also confirm the
relative performance of the three methods proposed in
(\ref{reliability_exact}), (\ref{square_reliability}), and
(\ref{mag_reliability}), denoted by $\Omega_{m}^{\textmd{Bayes}}$,
$\Omega_{m}^{|\cdot|,\theta}$, and $\Omega_{m}^{|\cdot|}$,
respectively, as well as the reversed relative performance of the
\emph{least} reliable tone set of each, which we generically denote
by $\arg\left\{\mathfrak{R}_{i:N}\right\}_{i=1}^{m}$.

\begin{figure}
\centering
\includegraphics[width=3.5in]{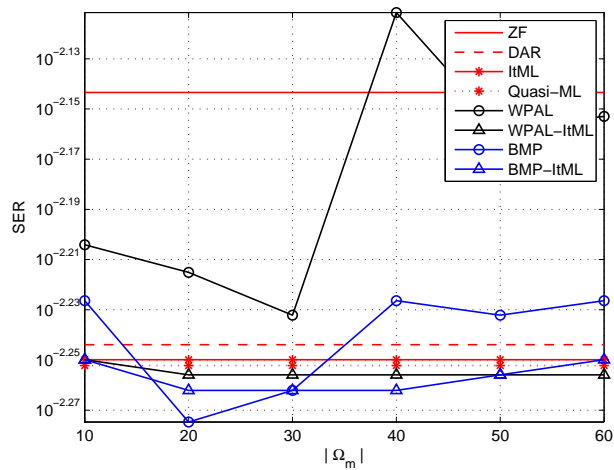}
\caption{Performance Comparison of CS techniques (alone and over
ItML) with  ItML \cite{ItML}, DAR \cite{DAR}, and Quasi-ML
\cite{Quasi} as a varying amount of the most reliable observations
in $\Omega_{m}^{|\cdot|,\theta}$ are considered.} \label{SER_vs_m}
\end{figure}

Furthermore, using our practical reliability function
(\ref{square_reliability}) based on (\ref{g_function}), we compared
our results with what we consider the most popular nonlinear
distortion mitigation techniques in the literature, namely, the
Iterative ML Decoding (ItML)\cite{ItML} and the Decision-Aided
Reconstruction (DAR)\cite{DAR} techniques. In addition, we also
implemented the Quasi-ML technique in \cite{Quasi} which proposed
improving the algorithm in \cite{ItML} by refraining from making
hard decisions when the absolute value of the real or imaginary part
of the frequency deviation is larger than some linear function
$\epsilon$ of $d_{min}$. Results in Fig. \ref{SER_vs_m} show the
superior performance of using BMP \cite{Schniter2} over the set
$\Omega_{m}^{|\cdot|,\theta}$, using only half the tones to reach
the optimum performance. The WPAL performs significantly better than
Zero Forcing (ZF), and can be used to improve the results of ItML,
even though it performs less efficiently alone under most
circumstances. Lastly, no gain is achieved by supplying the BMP
estimate to ItML, as BMP alone normally outperforms this procedure.

\section{Conclusion}
A novel method has been proposed to use data-aided CS techniques
over a reliable subset of observations in the frequency domain in
order to estimate and cancel sparse nonlinear distortion on an OFDM
signal in the time domain. Moreover, a newly developed method of
computing the reliability of each observation independently of the
other $M-1$ candidates within a constellation was also proposed and
tested. The methods offer promising performance, and the authors are
considering several possible improvements such as invoking soft
decoding and CNR maximization.



%


%
%

\ifCLASSOPTIONcaptionsoff
  \newpage
\fi




\bibliographystyle{IEEEtran}


%


\bibliographystyle{plain}

\newpage

%








\end{document}